\documentclass[showpacs,twocolumn,amsmath,amssymb,prl]{revtex4}

\usepackage{graphicx}
\usepackage{bm}
\usepackage{subfigure}
\usepackage{overpic}
\usepackage{color}

\newcommand{\ua}{\uparrow}
\newcommand{\da}{\downarrow}

\renewcommand{\epsilon}{\varepsilon}

\begin{document}

\title{
\bf\Large{Superconductivity with Finite-Momentum Pairing\\ in Zero Magnetic Field}}

\author{Florian Loder}
\author{Arno P. Kampf}
\author{Thilo Kopp}
\affiliation{
Center for Electronic Correlations and Magnetism, Institute of Physics, University of 
Augsburg, D-86135 Augsburg, Germany}

\begin{abstract}
In the BCS theory of superconductivity, one assumes that all Cooper pairs have the same center of mass momentum. This is indeed enforced by self consistency, if the pairing interaction is momentum independent. Here, we show that for an attractive nearest neighbor interaction, this is different. In this case, stable solutions with pairs with momenta ${\bf q}$ and $-{\bf q}$ coexist and, for a sufficiently strong interaction, one of these states becomes the groundstate of the superconductor. This finite-momentum pairing state is accompanied by a charge order with wave vector $2{\bf q}$. For a weak pairing interaction, the groundstate is a $d$-wave superconductor.
\end{abstract}


\date{\today}

\maketitle

In the original formulation of the BCS theory of superconductivity \cite{bcs}, all Cooper pairs are assumed to have the same center of mass momentum ${\bf q}$. One possible generalization of this theory is to introduce a pair amplitude for each center of mass momentum separately. In BCS theory for conventional superconductors, only one of these order parameters (OPs) is selected and the stable state is the one where all pairs have the same momentum and form the BCS condensate. For films in an external magnetic field, Fulde and Ferrell and, independently, Larkin and Ovchinnikov \cite{fulde:64, larkin:64} introduced a superconducting (SC) state with coexisting pair momenta $\pm{\bf q}$, a state that explicitly breaks time inversion symmetry.
For unconventional pairing symmetries, the competition between pair momenta is more complex and it has remained unresolved whether a bulk SC groundstate with different pair momenta may exist without magnetic field \cite{vorontsov:09}.

A SC state with different coexisting pair momenta generally exhibits a spatially inhomogeneous charge density. One example of a superconductor of this type is the recently proposed \textquotedblleft pair density wave\textquotedblright (PDW) state \cite{agterberg:08, baruch:08, berg:09}. It is characterized in real space by a two-component order parameter $\Delta({\bf r})=\Delta_{\bf q} e^{i{\bf q}{\bf r}}+\Delta_{-{\bf q}}e^{-i{\bf q}{\bf r}}$. This structure bears some resemblance to the Larkin-Ovchinnikov state, but it preserves time inversion symmetry. The PDW is accompanied by a charge density pattern with wave vector $2{\bf q}$. For this reason the PDW state has been proposed to describe the SC state of high-$T_c$ cuprates with coexisting stripe order, especially Nd-doped La$_{2-x}$Sr$_x$CuO$_4$ \cite{tranquada:95} and La$_{2-x}$Ba$_x$CuO$_4$ for $x=1/8$ \cite{fujita:04, abbamonte:05, berg:07, tranquada:08}. In particular, the recent experiments on the $1/8$-doped material stimulated further theoretical studies to resolve the nature and the origin of the SC state in the charge ordered phase \cite{white:09}. The PDW might be a candidate state, but so far a microscopic model which yields the PDW as its groundstate is lacking.

In this article, we formulate an extended version of BCS theory using Gor'kov's equations for the Green's functions in the SC state. We explicitly allow for the coexistence of different finite-momentum pairing amplitudes in the absence of an external magnetic field. We identify conditions for a groundstate solution with finite OPs for the pair momenta $\pm{\bf q}$. This pairing state is realized beyond a critical interaction strength $V_c$ for an attractive nearest-neighbor interaction, and it is characterized by a charge stripe order, a gapless density of states (DOS) and a partially reconstructed Fermi surface.
On the other hand, for $V<V_c$, the $d$-wave superconductor is the stable groundstate.

We start from a tight binding Hamiltonian on a square lattice with $N$ sites and periodic boundary conditions
\begin{multline}
{\cal H}=\sum_{{\bf k},s}\epsilon_{\bf k} c^\dag_{{\bf k} s} c_{{\bf k} s}\ +\\\frac{1}{N}\sum_{\bf q}\sum_{{\bf k},{\bf k}'}\sum_{s,s'}V({\bf k},{\bf k}',{\bf q})c^\dag_{{\bf k} s} c^\dag_{-{\bf k}+{\bf q} s'}c_{-{\bf k}'+{\bf q} s'}c_{{\bf k}' s}.
\label{G1}
\end{multline}
With a nearest and next-nearest neighbor hopping amplitude $t$ and $t'$, respectively, the single-electron dispersion has the form
\begin{align}
\epsilon_{\bf k}=-2t\left[\cos k_x+\cos k_y\right] +4t'\cos k_x\cos k_y-\mu
\label{G2}
\end{align}
where $\mu$ is the chemical potential.

For the superconducting state with singlet pairing, we use the BCS type mean-field decoupling scheme and approximate $\langle c^\dag_{{\bf k}\ua}c^\dag_{-{\bf k}+{\bf q}\da}c_{-{\bf k}'+{\bf q}\da}c_{{\bf k}'\ua}\rangle\rightarrow\langle c^\dag_{{\bf k}\ua}c^\dag_{-{\bf k}+{\bf q}\da}\rangle\langle c_{-{\bf k}'+{\bf q}\da}c_{{\bf k}'\ua}\rangle$. The system is then represented by the spin independent imaginary time Green's function ${\cal G}({\bf k},{\bf k}',\tau)=-\langle T_\tau c_{{\bf k} s}(\tau)c^\dag_{{\bf k}'s}(0)\rangle$, and the anomalous propagators ${\cal F}({\bf k},{\bf k}',\tau)=\langle T_\tau c_{{\bf k} s}(\tau)c_{-{\bf k}'s'}(0)\rangle$ and ${\cal F}^*({\bf k},{\bf k}',\tau)=\langle T_\tau  c^\dag_{-{\bf k} s}(\tau)c^\dag_{{\bf k}'s'}(0)\rangle$ for $s\neq s'$. The Heisenberg equations of motion for the normal and anomalous Green's functions lead to the Gor'kov equations \cite{agd7}:
\begin{widetext}\vspace{-5mm} 
\begin{align}
{\cal G}({\bf k},{\bf k}',\omega_n)&={\cal G}_0({\bf k},\omega_n)\left[\delta_{{\bf k}{\bf k}'}-\sum_{\bf q}\Delta({\bf k},{\bf q}){\cal F}^*({\bf k}-{\bf q},{\bf k}',\omega_n)\right]^{-1},
\label{G5}\\
{\cal F}({\bf k},{\bf k}',\omega_n)&={\cal G}_0({\bf k},\omega_n)\sum_{\bf q}\Delta({\bf k},{\bf q}){\cal G}(-{\bf k}',-{\bf k}+{\bf q},-\omega_n),
\label{G6}
\end{align}
\end{widetext}
where ${\cal G}_0({\bf k},\omega_n)=\left[i\omega_n-\epsilon_{\bf k}\right]^{-1}$ is the Green's function in the normal state and $\omega_n=(2n-1)\pi T$ is the fermion Matsubara frequency for temperature $T$. The order parameter $\Delta({\bf k},{\bf q})$ is determined by the self-consisteny condition
\begin{align}
\Delta({\bf k},{\bf q})&=-\frac{T}{N}\sum_n\sum_{{\bf k}'}V({\bf k},{\bf k}',{\bf q}){\cal F}({\bf k}',{\bf k}'-{\bf q},\omega_n).
\label{G3}
\end{align}
For the interaction, we choose a simple ansatz that allows for unconventional pairing; we assume an attractive interaction between electrons on neighboring sites.
In the singlet channel this is equivalent to the interaction $V({\bf k},{\bf k}',{\bf q})=V_s({\bf k},{\bf k}',{\bf q})+V_d({\bf k},{\bf k}',{\bf q})$ in momentum space, with factorizable extended $s$-wave and $d$-wave components $V_s({\bf k},{\bf k}',{\bf q})$ and $V_d({\bf k},{\bf k}',{\bf q})$, where
\begin{align}
V_{s,d}({\bf k},{\bf k}',{\bf q})=Vg_{s,d}({\bf k}-{\bf q}/2)g_{s,d}({\bf k}'-{\bf q}/2).
\label{G4}
\end{align}
$V>0$ is the attractive pairing interaction strength and $g_s({\bf k})=\cos k_x+\cos k_y$ and $g_d({\bf k})=\cos k_x-\cos k_y$. Thus 
\begin{align}
\Delta({\bf k},{\bf q})=\Delta_s({\bf q})g_s({\bf k}-{\bf q}/2)+\Delta_d({\bf q})g_d({\bf k}-{\bf q}/2).
\label{G4.1}
\end{align}
The vector ${\bf q}$ labels mean-field solutions which correspond to order parameters in real space with phase winding numbers $q_x$ and $q_y$ in $x$- and $y$-direction, respectively.

If $\Delta({\bf k},{\bf q})\neq\bm0$ for a single momentum ${\bf q}\neq\bm0$, then ${\cal F}({\bf k},{\bf k}',\omega_n)$ and ${\cal F}^*({\bf k},{\bf k}',\omega_n)$ have off-diagonal terms in momentum space, but ${\cal G}({\bf k},{\bf k}',\omega_n)$ is still diagonal. If $\Delta({\bf k},{\bf q})\neq0$ for at least two different momenta ${\bf q}$, then also ${\cal G}({\bf k},{\bf k}',\omega_n)$ has off-diagonal terms and the discrete translational invariance is broken. The charge density is obtained from $\rho({\bf r})=1/N\sum_{{\bf k},{\bf k}'} e^{i{\bf r}\cdot({\bf k}-{\bf k}')}n({\bf k},{\bf k}')$, where $n({\bf k},{\bf k}')=2T\sum_n{\cal G}({\bf k},{\bf k}',\omega_n)$. Thus there are charge modulations whenever ${\cal G}({\bf k},{\bf k}',\omega_n)$ has off-diagonal terms.

Inserting Eq.~(\ref{G6}) into Eq.~(\ref{G5}) leads to a system of coupled equations for the Green's function ${\cal G}({\bf k},{\bf k}',\omega_n)$. Assuming that $n({\bf k},{\bf k}')\ll n({\bf k})$ for ${\bf k}\neq{\bf k}'$, which will be verified a posteriori, ${\cal F}({\bf k},{\bf k}+{\bf q},\omega_n)$ and ${\cal F}^*({\bf k},{\bf k}+{\bf q},\omega_n)$ are approximated by keeping only the term proportional to ${\cal G}({\bf k}+{\bf q},{\bf k}+{\bf q},\omega_n)$ in Eq.~(\ref{G6}) and by neglecting the sum over ${\bf q}$ otherwise. Within this approximation the Gor'kov equations are solved analytically, as shown below.

In an ansatz for a self consistent solution of the Gor'kov equations~(\ref{G5}) and~(\ref{G6}), we choose $Q$ trial vectors ${\bf q}_1,\dots,{\bf q}_Q$ and set $\Delta({\bf k},{\bf q})=0$ for all other values of ${\bf q}\neq{\bf q}_i$. Thereby we test selected combinations of ${\bf q}$-vectors for self consistent solutions. With this ansatz, the energy spectrum of the system consists of $Q+1$ bands $E_\alpha({\bf k})$, $\alpha=0,\dots,Q$. The conventional BCS solution is realized for $Q=1$ with just two quasiparticle bands and ${\bf q}=\bm0$. Generally, one obtains a set of $2Q$ coupled self-consistency equations for $\Delta_s({\bf q}_i)$ and $\Delta_d({\bf q}_i)$:
\begin{align}
\frac{\Delta_{s,d}({\bf q})}{V}=-\frac{T}{N}\sum_{{\bf k}'}g_{s,d}({\bf k}'-{\bf q}_i/2)\sum_n{\cal F}({\bf k},{\bf k}-{\bf q}_i,\omega_n).
\label{G13}
\end{align}

Standard BCS theory provides isotropic solutions of Eq.~(\ref{G13}) with $\Delta_s(\bm0)=0$ and $\Delta_d(\bm0)\neq0$ for arbitrarily weak interaction strength $V$. Most remarkably, if $V$ exceeds a certain interaction strength, we identify solutions with $Q=2$ and with the specific set of ${\bf q}$-vectors $\{(q, 0), (-q, 0)\}$. These solutions are anisotropic and have a subdominant extended $s$-wave contribution $\Delta_s({\bf q})$, which increases with increasing $q$. Below we will discuss in particular the time inversion symmetric zero-current solutions of Eq.~(\ref{G13}), i.e., $\Delta_{s,d}({\bf q})=\Delta_{s,d}(-{\bf q})$.

To test the stability of this solution, we solved Eq.~(\ref{G13}) iteratively for selected combinations of ${\bf q}$ vectors and different initial values of the corresponding OPs. In particular, we investigated the stability of the above solution against decay into the ${\bf q}=\bm0$ state by using the ansatz with the three center of mass momenta $\{(q, 0), (-q, 0), (0, 0)\}$. We find that for a pure on-site interaction $V({\bf k},{\bf k}',{\bf q})\equiv V_0$ ($s$-wave pairing), finite momentum pairing is unstable. All OPs with different ${\bf q}$ compete, even the ones with ${\bf q}$ and $-{\bf q}$. Thus states with $Q\geq2$ will always decay into a state with only one finite order parameter for on-site $s$-wave pairing. However, 
for the nearest-neighbor interaction~(\ref{G4}), additional stable 
solutions emerge. The finite momentum pairing solutions are typically stable for a wide range of $q$ values. Here the OPs for $\pm q$ do not compete, but rather support each other. The range of stability however decreases with decreasing $V$, and eventually disappears.

\begin{figure}[t]
\centering
\vspace{2mm}
\begin{overpic}
[width=0.99\columnwidth]{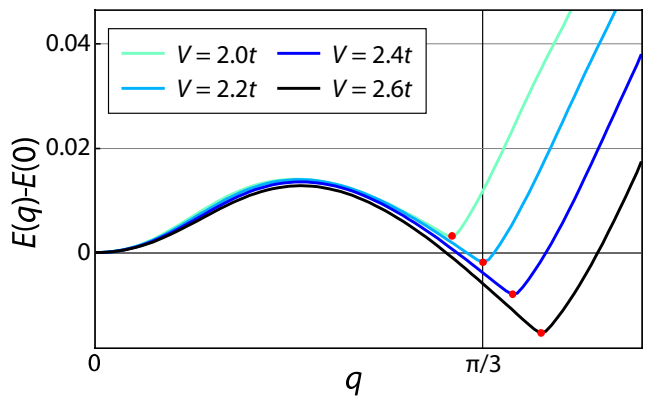}
\end{overpic}
\vspace{-6mm}
\caption{Energy $E=\langle{\cal H}\rangle$ as a function of pair momentum ${\bf q}=(q,0)$ for different pairing interaction strengths $V$. Calculations were performed for a $384\times384$ lattice with fixed electron density $\rho=0.8$ and $t'=0.3t$. For these parameters, the finite momentum pairing state becomes the groundstate for $V>V_c\approx2.2t$ with $q=\pi/3$.}
\label{FigG1}
\end{figure}

So far we have verified that stable finite momentum pairing solutions of the self-consistency equation exist. They refer to local minima of the free energy. To determine the groundstate at $T=0$, the global minimum of the energy $E=\langle{\cal H}\rangle=\sum_{\bf k} \epsilon_{\bf k} n({\bf k})+\sum_i[\Delta_s^2({\bf q}_i)+\Delta_d^2({\bf q}_i)]/V$ has to be determined with respect to all $q$. Figure~\ref{FigG1} shows the typical $q$-dependence of $E$ with a minimum at $q=0$ and a further minimum for $q>0$. The minimum at $q=0$ corresponds to the standard $d$-wave SC state. With increasing $V$, the energy of the minimum at finite $q$ decreases accompanied by a shift to larger $q$. 
This implies the existence of a critical interaction strength $V_c $, which depends on $t'$ and the electron density $\rho=1/N\sum_{\bf k} n({\bf k})$. Above $V_c$ the finite momentum pairing state is the groundstate. The optimal $q$ sensitively depends on $V$, $t'$ and $\rho$, but it is typically found in between $q\approx\pi/8$ and $q\approx\pi/2$ for a wide parameter range. For even larger values of $V$, our calculations identified the groundstate to be of a checkerboard type with $Q=4$ and with pair momenta $\{(q,0),(-q,0),(0,q),(0,-q)\}$.

For the finite-$q$ groundstate solutions the charge density $\rho({\bf r})$ has an oscillatory part arising from the off-diagonal terms of the Green's function. For the ${\bf q}=(\pm q,0)$ state the charge density forms a sinusoidal stripe pattern with wave number $2q$. Correspondingly, the charge density varies as
\begin{align}
\rho({\bf r})=\rho+\rho_1\cos(2qx),
\label{G15}
\end{align}
with an amplitude $\rho_1/\rho\approx2\%$, which justifies the assumption of small charge modulations in the above approximation for ${\cal F}({\bf k},{\bf k}',\omega_n)$. For $q=\pi/3$, the wavelength of the stripe pattern is therefore three lattice constants. The charge modulation in the SC state suggests to include a self-consistent charge density-wave (CDW) OP in the mean-field decoupling scheme of the Hamiltonian~(\ref{G1}). We have analyzed this  extension with coexisting OPs for SC and CDW order for selected cases. The CDW OP tends to stabilize the state with finite momentum pairing but it remains small and does not change the solutions qualitatively.

\begin{figure}[t]
\centering
\vspace{2mm}
\begin{overpic}
[width=0.99\columnwidth]{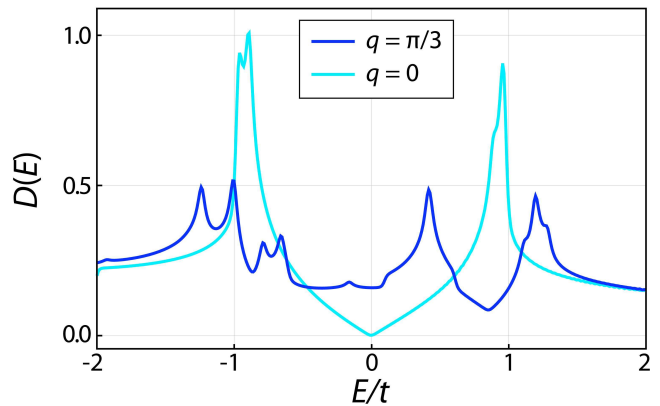}
\end{overpic}
\vspace{-6mm}
\caption{Density of states $D(E)$ of the groundstate solutions for interaction strength $V=2t$ and $V=2.2t$, corresponding to $q=0$ and $q=\pi/3$, respectively. The other parameters are the same as in Fig.~\ref{FigG1}.
The coherence peaks of the ${\bf q}=\bm0$ state are split due to the van Hove singularity of the 2D tight-binding dispersion.}
\label{FigG2}
\vspace{-3mm}
\end{figure}

\begin{figure}[t]
\centering
\vspace{6mm}
\begin{overpic}
[width=0.48\columnwidth]{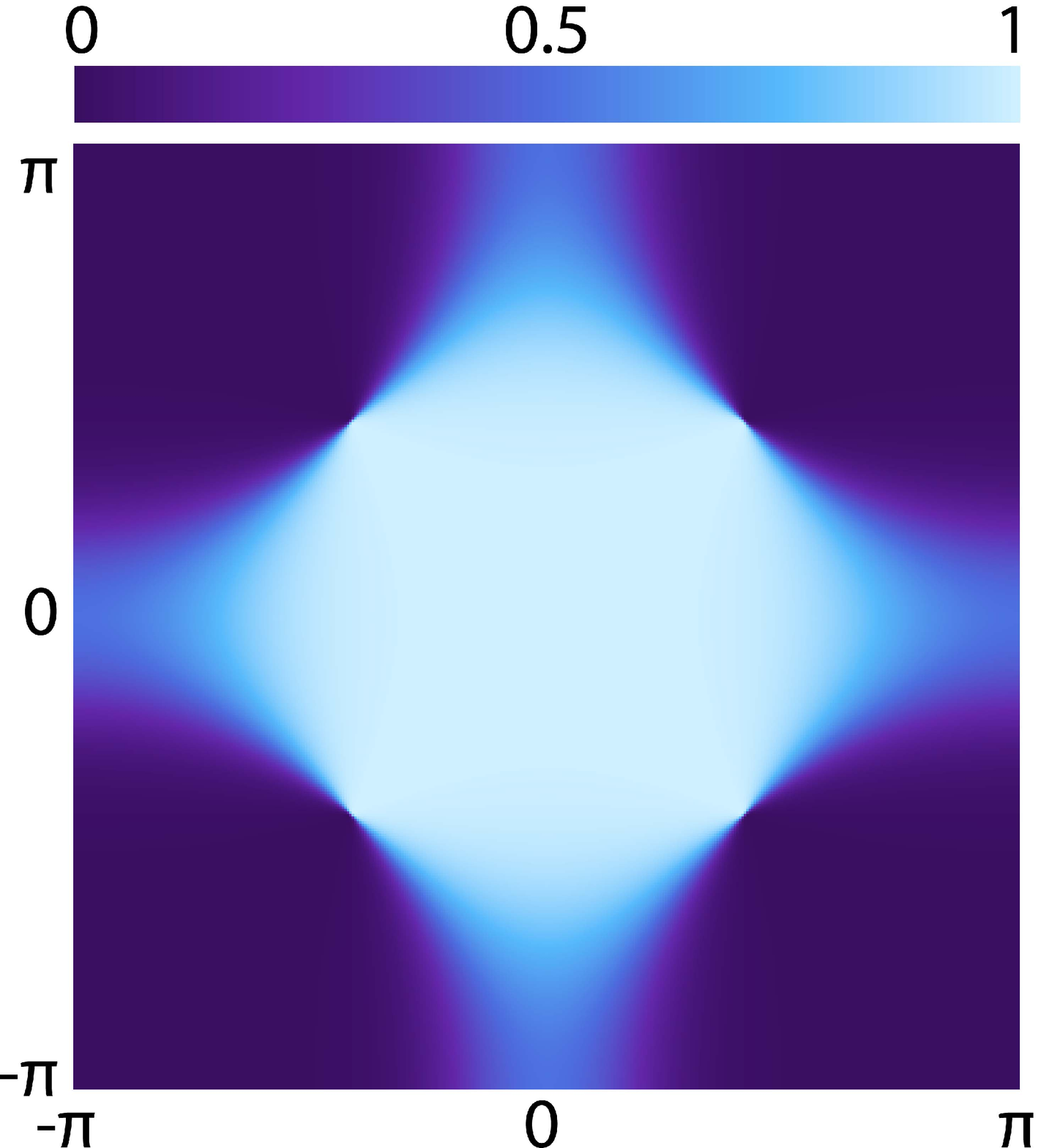}
\put(0,103){\bf(a)}
\end{overpic}
\begin{overpic}
[width=0.48\columnwidth]{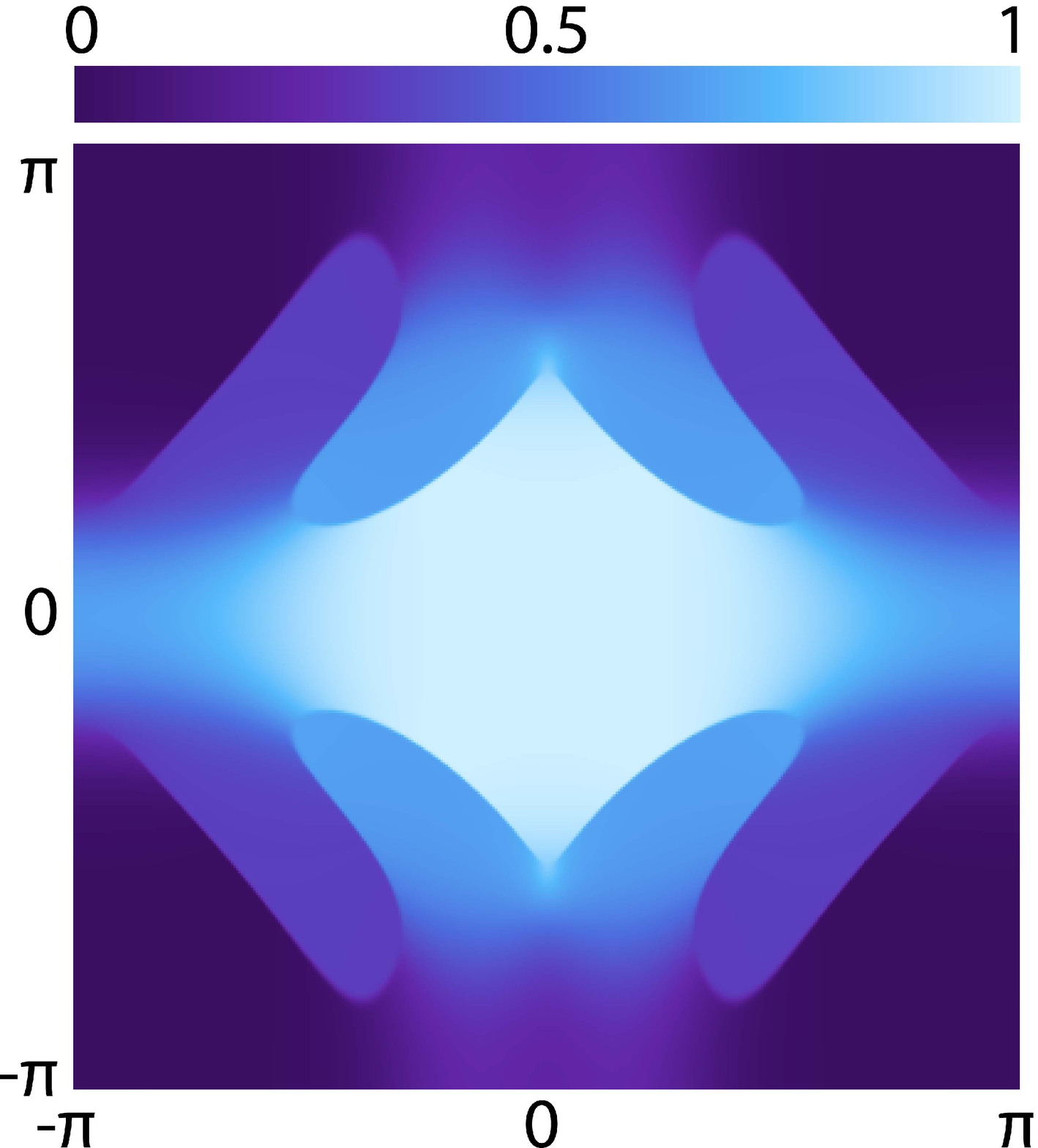}
\put(0,103){\bf(b)}
\end{overpic}\\[6mm]
\begin{overpic}
[width=0.48\columnwidth]{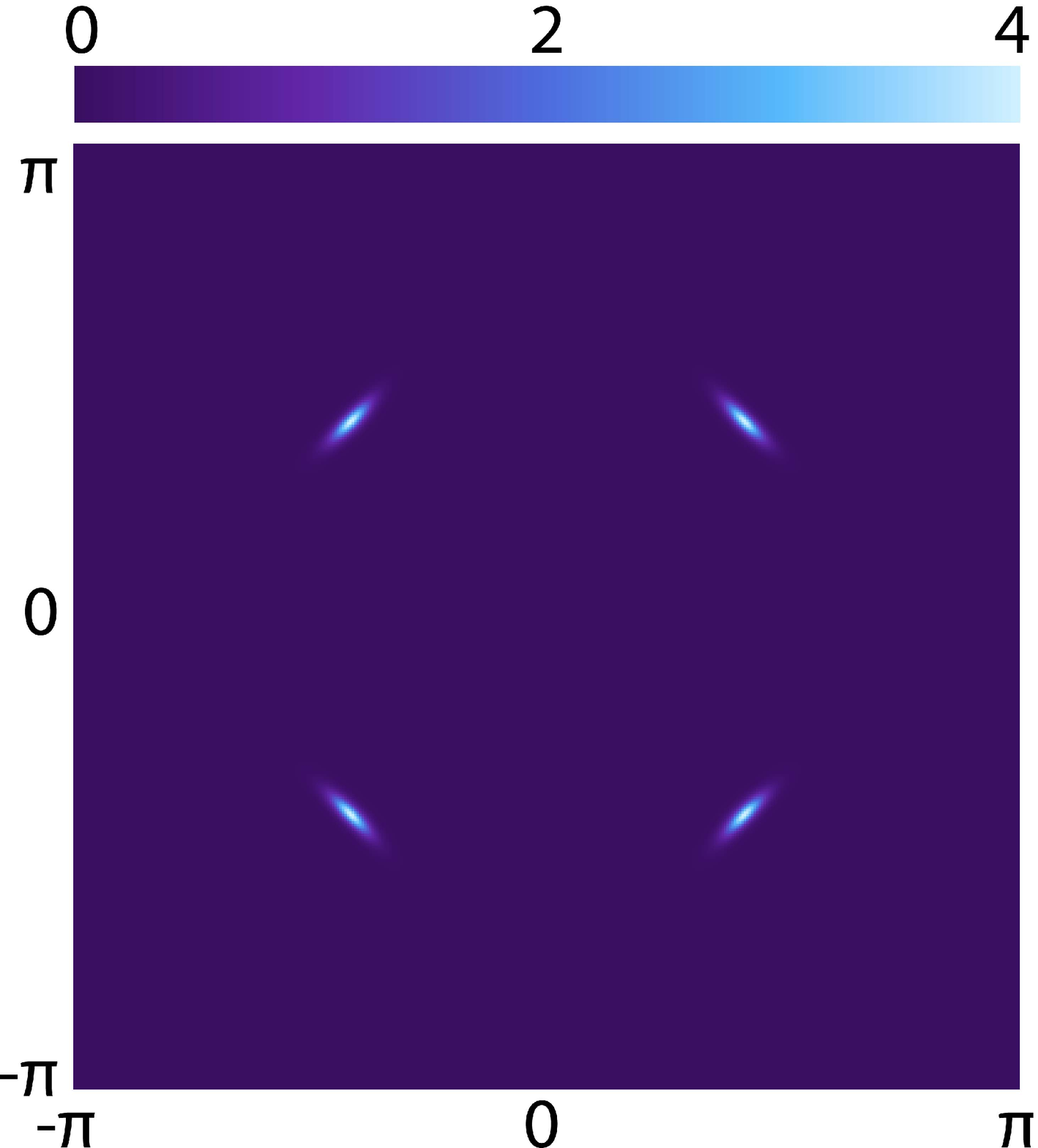}
\put(0,103){\bf(c)}
\end{overpic}
\begin{overpic}
[width=0.48\columnwidth]{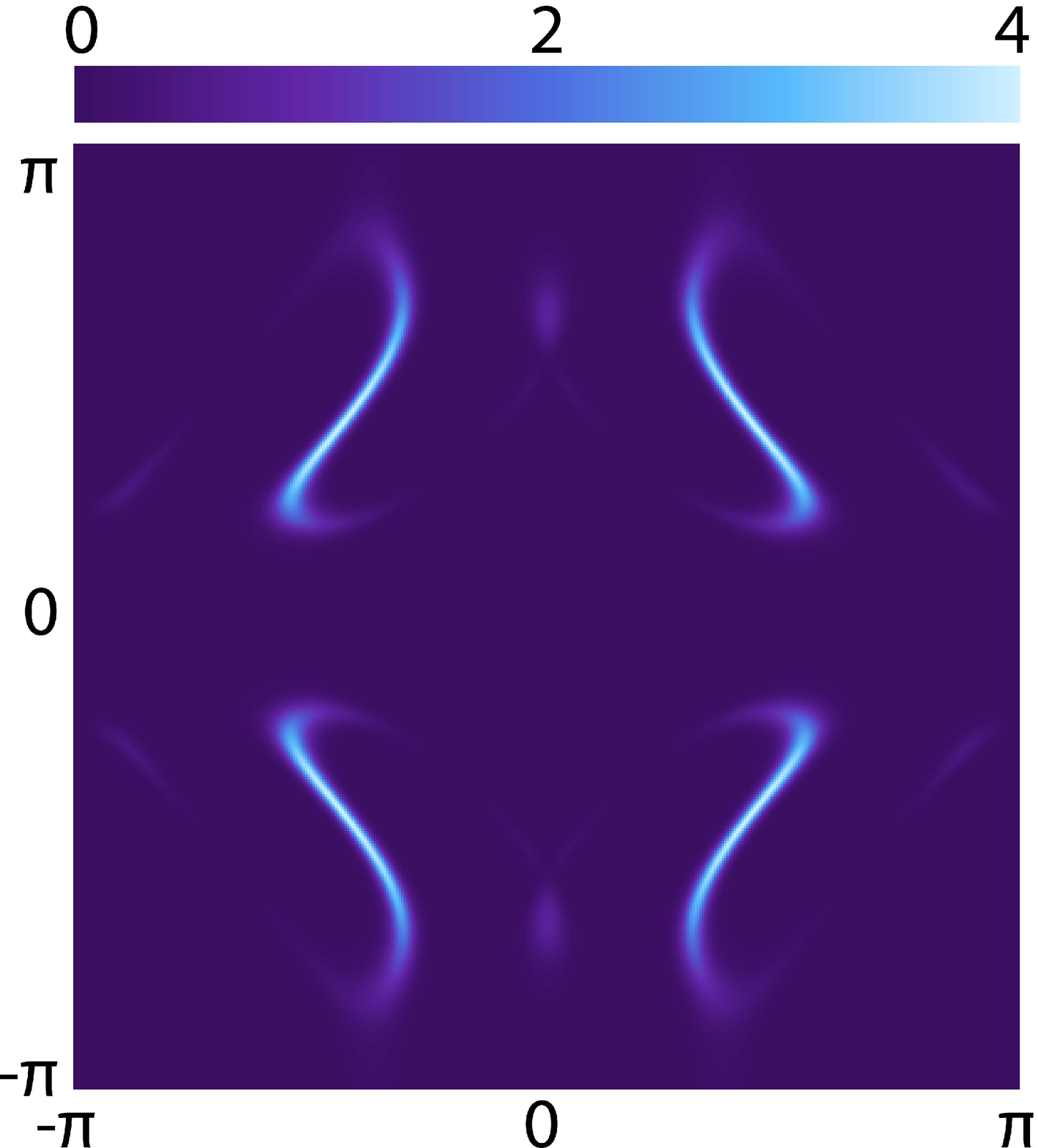}
\put(0,103){\bf(d)}
\end{overpic}\\[6mm]
\begin{overpic}
[width=0.48\columnwidth]{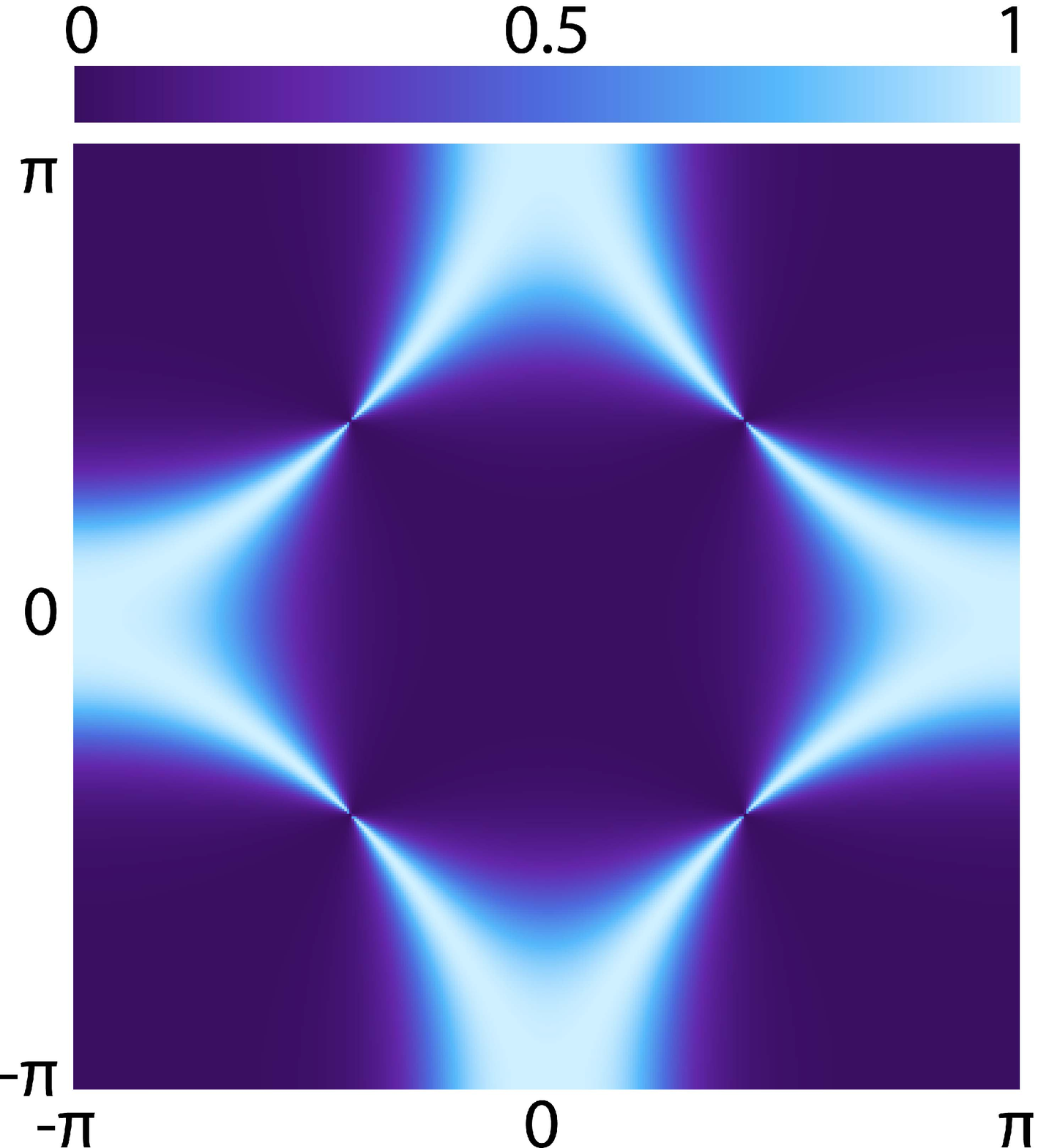}
\put(0,103){\bf(e)}
\end{overpic}
\begin{overpic}
[width=0.48\columnwidth]{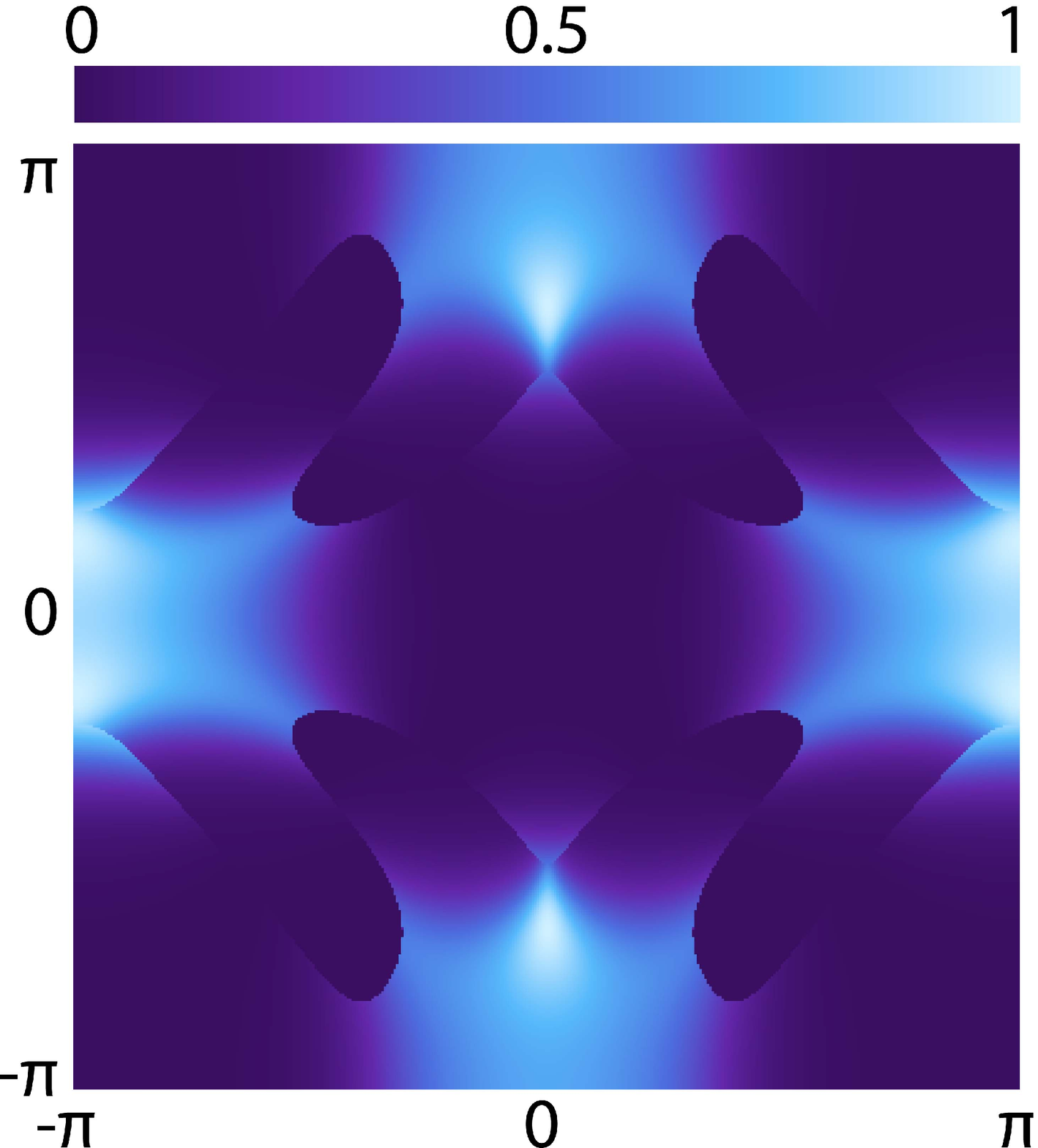}
\put(0,103){\bf(f)}
\end{overpic}
\caption{Momentum space properties of the finite momentum pairing state with $q=\pi/3$ and the same parameters as in Fig.~\ref{FigG2} (right panels) and for comparison the $d$-wave superconductor for $q=0$ (left panels).
(a),$\,$(b): Occupation probability function $n({\bf k})$.
(c),$\,$(d): Density of states with zero energy ${\rm Im\,} G({\bf k},{\bf k},0-i\delta)$ (here: $\delta=0.04t$).
(e),$\,$(f): Pair density $P({\bf k})$.}
\label{FigG3}
\vspace{-3mm}
\end{figure}

The finite momentum pairing state has further characteristic properties that are at variance with a BCS like $d$-wave superconductor (with $q=0$). The DOS $D(E)=\sum_{\bf k} {\rm Im\,} G({\bf k},{\bf k},E-i0^+)$, where ${\rm Im\,} G$ is the imaginary part of the analytical continuation of ${\cal G}$ to the real frequency axis, is shown in Fig.~\ref{FigG2}. For $q=\pi/3$, the DOS bears little resemblance to a $d$-wave like gap as the coherence peaks are split and the DOS is finite at the Fermi energy. A similar splitting is observed for current carrying $d$-wave states \cite{khavkine:04,loder:09} which originates from the Doppler shift of the finite momentum eigenstates.

Figure~\ref{FigG3} displays the characteristic momentum space properties of the finite momentum pairing state and, for comparison, of the $q=0$ $d$-wave superconductor. In the finite momentum pairing state, the momentum distribution function $n({\bf k})=n({\bf k},{\bf k})$ develops structures with sharp boundaries. These boundaries consist of lines in momentum space with $E_\alpha({\bf k})=0$, for $\alpha=0,1$ or 2 
indicative of a Fermi surface reconstruction. The zero energy states generate Fermi-arc like structures as shown in Fig.~\ref{FigG3}d.  For $t'=0$, $n({\bf k})$ is similar to the result obtained in Ref.~\cite{baruch:08}. The pair density $P({\bf k})=\sum_iP({\bf k},{\bf k}-{\bf q}_i)$, where $P^2({\bf k},{\bf k}')=2\langle c^\dag_{{\bf k}\ua}c^\dag_{-{\bf k}'\da}\rangle\langle c_{-{\bf k}'\da}c_{{\bf k}\ua}\rangle$, clearly shows the fingershaped ${\bf k}$-space structures of $n({\bf k})$ contain unpaired electrons only. The overall number of pairs is smaller in the finite momentum pairing state than in the $q=0$ state. This seems to contradict the fact that it has the lower energy. The latter, however, consists of both the kinetic energy, which rises in the SC state and acts against the formation of pairs, and the gain of condensation energy. The optimal balance between these two contributions depends on details of the single particle kinetic energy $\epsilon_{\bf k}$ and the interaction potential $V({\bf k},{\bf k}',{\bf q})$ and does not generally favor a larger number of paired electrons.

In this article we have shown that the extended BCS theory with attractive nearest neighbor interaction provides self consistent solutions with the simultaneous formation of electron pairs with center of mass momenta ${\bf q}$ and $-{\bf q}$. It is a microscopic solution which constitutes a stable macroscopic state of the PDW type which was proposed to describe the striped SC phase in hole doped 214 cuprates. This finite momentum pairing state is the groundstate beyond a critical interaction strength $V_c$. $V_c$ depends sensitively on the band filling $\rho$ and ranges from $V_c\approx1.4t$ for $\rho=0.6$ to $V_c\approx3.5t$ for $\rho=1$. This is consistent with the result in Ref.~\cite{berg:09} that only the uniform phase with fixed ${\bf q}$ can be the groundstate of the BCS Hamiltonian in the weak coupling limit.

The results described above as solutions of Gor'kov's equations are alternatively obtained by diagonalizing the mean-filed decoupled Hamiltonian (\ref{G1}) using a Bogoliubov transformation or by solving an extended version of Bogoliubov -- de Gennes equations in real space \cite{loder:10}.

Our results demonstrate as a proof of principle that stable groundstate solutions of the pairing Hamiltonian (\ref{G1}) exist with coexisting finite momentum pairing amplitudes for center of mass momenta ${\bf q}=(q,0)$ and $-{\bf q}$; these solutions are absent for an attractive contact interaction. Due to the concomitant striped charge density modulation with wave vector $2{\bf q}$, a connection to the striped superconductor La$_{15/8}$Ba$_{1/8}$CuO$_4$ appears tempting. However, without the inclusion of additional correlation effects as the source for a possible spin order pattern, we consider it premature to draw conclusions about the favorable wavelength of the stripes.

This work was supported by the Deutsche Forschungsgemeinschaft through SFB 484. We acknowledge stimulating discussions with Raymond Fr\'esard, Jochen Mannhart and Dieter Vollhardt.

\end{document}